\documentclass[seceq]{ptptex}
\newcommand{\la}{\langle}
\newcommand{\ra}{\rangle}
\newcommand{\lla}{\langle\langle}
\newcommand{\rra}{\rangle\rangle}

\newcommand{\Tr}{{\rm Tr}}

\newcommand{\beq}{\begin{eqnarray}}
\newcommand{\eeq}{\end{eqnarray}}
\newcommand{\sbeq}{\begin{subeqnarray}}
\newcommand{\seeq}{\end{subeqnarray}}
\newcommand{\eps}{\epsilon}

\newcommand{\bfzero}{\mbox{{\boldmath $0$}}}

\newcommand{\bfm}{\mbox{{\boldmath $m$}}}
\newcommand{\bfp}{\mbox{{\boldmath $p$}}}

\newcommand{\bfx}{\mbox{{\boldmath $x$}}}

\newcommand{\bfD}{\mbox{{\boldmath $D$}}}

\newcommand{\bfG}{\mbox{{\boldmath $G$}}}

\newcommand{\bfM}{\mbox{{\boldmath $M$}}}

\newcommand{\bfq}{\mbox{{\boldmath $q$}}}

\newcommand{\bfT}{\mbox{{\boldmath $T$}}}

\newcommand{\btheta}{\mbox{{\boldmath $\theta$}}}

\newcommand{\bfPi}{\mbox{{\boldmath $\Pi$}}}

\newcommand{\bfeta}{\mbox{{\boldmath $\eta$}}}

\title{%        %You can use \\ for explicit line-break.
 $X$ Meson aka $\eta'$ and Kobayashi-Maskawa-'t Hooft Six-quark Vertex%
}

\subtitle{$U(1)_A$ Anomaly and Generalized Nambu-Jona-Lasinio Model}
\author{%       %Use \scshape for the family name.
Teiji \textsc{Kunihiro}%
}

\inst{%     %Affiliation, neglected when [addenda] or [errata].
Department of Physics, Kyoto University,
Kitashirakawa, Kyoto 606-8502, Japan
}

\abst{%         %This abstract is neglected when [addenda] or [errata].
In  1970, Kobayashi and Maskawa concluded that 
an effective  six-quark vertex with a determinantal form is necessary
in the chiral effective models to
 account for the large mass of $X$ meson,  
which is now called the $\eta'$.
The determinantal interaction  has an $SU(3)_L \otimes SU(3)_R$ 
symmetry but
not $U(3)_L \otimes U(3)_R$, and, hence accounts for 
the explicit breaking of $U(1)_A$ symmetry in quantum chromodynamics (QCD);
the vertex was later derived by 't Hooft as an instanton-induced
quark interaction. The vertex, which may be called the Kobayashi-Maskawa-'t Hooft
(KMT) term, is widely used in quantitative analyses of
hadron physics and QCD phase transitions at finite temperature
and density. An account is made for the
KMT term with recent extensive applications.  Described are also
personal experiences
with Professor Maskawa and Professor Kobayashi,
including an encounter with Professor Maskawa when the author 
first presented his work on 
 the KMT term.
}

\begin{document}

\maketitle

\section{Introduction}

The pseudoscalar meson $\eta'(958)$  was called  $X$ in the 
past for some time.
It was difficult to understand its large mass within 
the chiral $SU(3)_L \otimes SU(3)_R$ with an explicit symmetry breaking 
term $-\eps_0 S_0 -\eps_8 S_8$, which may have been  identified with the 
quark mass term with $S_a=\bar{q}\lambda_a q$; here, the isospin symmetry is 
assumed.  Note that the existence of the quarks 
or {\em ur}-baryons 
was far from being established in those days.
As early as  1970, 
Kobayashi and Maskawa
indicated, in a paper\cite{Kobayashi:1970ji} entitled ``Chiral symmetry and 
eta-$X$ mixing'', 
that this is a serious problem, 
%Partly recourse to Nambu-Jona-Lasinio model,
and concluded  that there should exist a six-quark interaction with 
a determinantal form
\beq
\det_{i,j}\bar{q}_i (1-\gamma_5) q_j + {\rm h.c.},
\label{eq:kmt}
\eeq
where h.c. stands for Hermite conjugate.
Their analysis was based on 
 the method of Gell-Mann-Oakes-Renner\cite{GellMann:1968rz},
and a detailed account of the outcome of this vertex was reported
in 1971 by Kobayashi, Kondo and Maskawa\cite{Kobayashi:1971qz}.
This vertex is contained in instanton-induced quark interaction derived
by 't Hooft in 1976\cite{'tHooft:1976fv}, 
and is often called the 't Hooft vertex.
However, one now  sees that the appropriate name of this six-fermion
determinantal vertex should be the Kobayashi-Maskawa or
at least Kobayashi-Maskawa-'t Hooft term, which we adopt\cite{Hatsuda:1994pi}
 and will be abbreviated as the KMT term or vertex.

The compatibility of the large mass of the $\eta'$ with quantum chromodynamics
 (QCD) was 
formulated as the $U(1)_A$ problem by  Weinberg\cite{Weinberg:1975ui} in 
1975.
The present understanding of the resolution of the problem is 
also described in the textbook by Weinberg\cite{Weinberg:1996kr}
 and also by 
Fujikawa and Suzuki\cite{Fujikawa:2004cx}, for example.
 The basic ingredients are the $U(1)_A$ anomaly
for the divergence of the axial current in the flavor singlet
and the instanton configuration\cite{Belavin:1975fg} leading to the
$\theta$-vacuum\cite{Jackiw:1976pf}.
The physical origin of
 the large mass of the $\eta'$
is thus understood to be due to 
an explicit breaking of 
the  $U(1)_A$ symmetry. It means that
the low-energy effective theory of QCD
should contain a vertex that explicitly  breaks
the $U(1)_A$ symmetry.

An interesting point of the work by Kobayashi and Maskawa
is the fact that their proposal was
based on the work by Nambu\cite{Nambu:1961tp}
 who shared the Nobel prize with them in 2008.
Noting that 
the original Nambu-Jona-Lasinio (NJL) model\cite{Nambu:1961tp} only contains
the four-fermion interaction and, hence, becomes inevitably 
$U(3)_L \otimes U(3)_R$
invariant even if one imposes $SU(3)_L \otimes SU(3)_R$
invariance to the model, Kobayashi and Maskawa\cite{Kobayashi:1970ji} concluded
that 
the six-fermion interaction of a determinantal form
should be present in the chiral quark model 
as given by Nambu-Jona-Lasinio\cite{Nambu:1961tp} for the $\eta'$ to be
described in the theory.\footnote{The same model
was later proposed independently by Mirelli and Schechter in 1976\cite{Mirelli:1976ww}.}

The first serious and extensive analyses of such an extended
NJL model with the six-fermion determinantal
interaction were carried out  around 1987 to 1988 by several people
including the present author\cite{soken88,Kunihiro:1987bb,Bernard:1987sg,Reinhardt:1988xu};
in these analyses,
the nonperturbative vacuum is determined
in the self-consistent mean-field theory,
 and the pseudoscalar and scalar mesons as collective excited states 
on top of the  vacuum are also calculated as in the
original work\cite{Nambu:1961tp}, and, hence, the model
parameters are determined  explicitly.
In those days, people became interested 
in the possible violation of the Okubo-Iizuka-Zweig rule in the baryon sector, 
which was prompted partly by the mysteriously large value
 of the $\pi$-$N$ sigma term $\Sigma_{\pi N}$\cite{Reya:1974gk},
which may be related to the possible strangeness content of the
nucleon in the scalar channel; see Ref. \citen{Ohki:2008ff} for a
 recent status of the understanding of this subject.
An interesting aspect of the KMT term is that it
can give rise to a flavor mixing in the scalar as well as 
in the pseudo-scalar channels as in the $\eta$-$\eta'$ system
and, hence, may be an origin of a possible
OZI rule violation in the baryon sector\cite{soken88,Bernard:1987sg,Kunihiro:1988vi}.
An extension of the model 
to include the vector and axial vector 
fields\cite{Reinhardt:1988xu,Takizawa:1989sv,Takizawa:1990ku,Klimt:1989pm}
is reviewed in Refs. \citen{Vogl:1991qt} and \citen{Klevansky:1992qe}.
The flavor mixing in the axial vector channel may be related to 
the ``Spin Crisis''\cite{Takizawa:1990ay,Takizawa:1991si}.

In this article, we will describe  the generalized
NJL model with the KMT term  and provide a
brief review on how it works 
as an effective theory of QCD, particularly in describing the
$\eta$-$\eta'$ system, and the scalar meson dynamics together with
other QCD phenomenology.
We  shall also provide a sketch on how the chiral quark 
models with the anomaly terms
are  utilized  to explore the properties of the quark/hadronic
matter  at finite temperature and/or density.
 The first part of the  following  will be
based on our previous review\cite{Hatsuda:1994pi};
see 
Ref.~\citen{Hatsuda:1994pi} for the details of  this part.\footnote{
Reference~\citen{Hatsuda:1994pi} was cited in his Nobel lecture by
Professor Y.~ Nambu, which was presented by Professor G.~Jona-Lasinio.}
One can also refer to Refs.~ \citen{Vogl:1991qt}, \citen{Klevansky:1992qe}
and \citen{Buballa:2003qv}
for a  complementary account of the subjects and 
useful references.
% and also the papers citing these reviews.
%It is my great pleasure and an honor that I have such  an opportunity
%to introduce the model proposed by  Kobayashi and Maskawa
%based on the work by Nambu, to which I might have made
%some contribution.
The author's personal experiences  with Professor Maskawa 
and Professor Kobayashi will  also be described, which
include an encounter when the author made 
the first presentation of his work on the generalized
Nambu-Jona-Lasinio model incorporating the KMT term in 1987.

\section{The NJL model with the KMT term}

\subsection{The model}

The effective model considered  by Kobayashi and Maskawa\cite{Kobayashi:1970ji}
was a generalization of the Nambu-Jona-Lasinio
 (NJL) model\cite{Nambu:1961tp} to the three-flavor case
 with the anomaly term incorporated
in the form of the determinantal interaction:
\beq
{\cal L} &=& \bar{q}i\gamma \cdot \partial q
+ \sum^{8}_{a=0}
{g_{_S} \over 2}[(\bar{q}\lambda_a q)^2 + (\bar{q}i\lambda_a
\gamma_5q)^2] -\bar {q}{\bfm}q+  g_{_{D}} [{\rm det}\bar{q}_i (1-\gamma_5) q_j
+{\rm h.c.}] \nonumber \\
 \ \ \ &\equiv& {\cal L}_0   +{\cal L}_S + {\cal L}_{SB} + {\cal L}_{_{KMT}} ,
\label{eq:KMNJL}
\eeq
where the quark field $q_i$ has three colors ($N_c=3$) and three flavors
($N_f=3$), $\lambda^a$ ($a = 0 \sim 8$) are the Gell-Mann matrices with
$\lambda_0$=$\sqrt{2 \over 3}\bf{1}$. 
Although not explicitly written in their paper,
this is exactly the model Lagrangian that Kobayashi and Maskawa
considered in their 1970 paper. Therefore, we call the Lagrangian
(\ref{eq:KMNJL}) the Kobayashi-Maskawa-Nambu-Jona-Lasinio (KM-NJL) model.
%although its serious analysis was first made in \cite{kunihiro87}.
 
We emphasize that the model embodies  three basic  ingredients of QCD, i.e.,
 the dynamical breaking of chiral symmetry (DBCS), $U(1)_A$ anomaly, and
the explicit symmetry breaking
 due to the current quark masses.
Extensive studies using this model showed\cite{Hatsuda:1994pi} that
 the  various empirical aspects of QCD are
 realized through the interplay among the three ingredients.
  It was  emphasized in Ref. \citen{Hatsuda:1994pi}
 that the constituent quark model and
 chiral symmetry are reconciled in a chiral quark model;
the chiral quark model can account for most of
  the empirical facts on baryons as well as the low-lying mesons.
 Furthermore,
such an effective model allows us to study the change in hadron
properties in hot/dense medium in a self-consistent manner, 
as initiated by T. Hatsuda and the present author 
for the two-flaver case \cite{HK85}.
Before entering into the physical consequences, let us examine its symmetry 
properties.

\subsection{Symmetry properties}

To observe the transformation properties of each term
in the KM-NJL model, it is convenient to introduce
  the 3$\times$3 bosonic matrices by
\beq
\Phi _{ij}=\bar{q}_j(1-\gamma _5)q_i=2\bar{q}_{_{jR}}q_{_{iL}}
=\bar{q}_jq_i\, +\, i\bar{q}_ji\gamma_5q_i,
\eeq
with
$(\Phi^{\dag}) _{ij}=\bar{q}_j(1 +\gamma _5)q_i=2\bar{q}_{_{jL}}q_{_{iR}}$, 
where
$q_{_{iL}}\equiv 1/2\cdot (1-\gamma _5)\, q_i$ and 
$q_{_{iR}}\equiv 1/2\cdot (1+\gamma _5)\, q_i$
are the left- and right-handed fields, respectively.
We  note that
\beq
\bar{q}_{_{\rm R}}\lambda_{a}q_{_{\rm L}}= {\rm Tr}[\lambda_{a}\Phi]/2
\equiv \Phi_a
\quad {\rm and}
\quad
\bar{q}_{_{\rm L}}\lambda_{a}q_{_{\rm R}}= {\rm Tr}[\lambda_a \Phi^{\dag}]/2
\equiv \Phi^{\dag}_a,
\eeq
and accordingly,
 $\bar{q}\lambda_aq=(\Phi^{\dag}_a +\Phi_a)/2$ and
$\bar{q}i\gamma_5\lambda_aq=i(\Phi^{\dag}_a -\Phi_a)/2$.
Then, the Lagrangian is cast into a form reminiscent of the linear
$\sigma$-model:
\beq
{\cal L}_s + {\cal L}_{SB}+{\cal L}_{_{KMT}} 
=g_s {\rm Tr}(\Phi^{\dagger} \Phi )
  - {1 \over 2}{\rm Tr}[\bfm (\Phi + \Phi^{\dagger})]
+g_D ({\rm det}\Phi + {\rm h.c.}).
\eeq

%Then, one sees that ${\cal L}_{_S}$ is rewritten as
%\beq
%{\cal L}_{S}=g_{_S}{\rm Tr}(\Phi^{\dag}\Phi).
%\eeq

Now the chiral $SU(3)_L \otimes SU(3)_R$ transformation is defined by
\beq
q_{_{iL}}\rightarrow [U(\btheta _{\rm L})]_{ij}q_{_{jL}}\equiv L_{ij}q_{_{jL}},\ \ 
q_{_{iR}}\rightarrow [U(\theta _{\rm R})]_{ij}q_{_{jR}}\equiv R_{ij}q_{_{jR}},
\eeq
with 
$U(\btheta)={\rm exp}(i\sum_{a=1 \sim 8}\theta_a\lambda_a/2)$,\,
the determinant of which is unity. Here, the repeated suffix
implies a summation over it.
 Under the $SU(3)_L \otimes SU(3)_R$ transformation, the bosonic 
operators are transformed as
\beq
\Phi_{ij}\rightarrow L_{ik}\Phi_{kl}
R^{\dag}_{lj},\ \ \ 
\Phi^{\dag}_{ij}\rightarrow R_{ik}\Phi^{\dag}_{kl}L^{\dag}_{lj},
\eeq
which shows that they are  representations $(3,\bar 3)$ and 
$(\bar 3, 3)$ of $SU(3)_L \otimes SU(3)_R$, respectively.
Thus, it is easily seen that ${\cal L}_S$ 
has an $SU(3)_L \otimes SU(3)_R$ invariance.
A notable point is that ${\cal L}_S$ is also {\em automatically}
invariant under
the $U(1)_L \otimes U(1)_R =\,U(1)_V \otimes U(1)_A$
 transformation defined by
\beq
q_{_{\rm L}} \rightarrow e^{i\alpha_0/2}q_{_{\rm L}},\ \ \ 
q_{_{\rm R}} \rightarrow e^{i\beta_0/2}q_{_{\rm R}}.
 \eeq
In fact, under this transformation, the bosonic variables are 
transformed as
\beq
\Phi \rightarrow e^{i(\alpha_0 -\beta_0)/2}\Phi,\ \ \ 
\Phi ^{\dag} \rightarrow e^{-i(\alpha_0 -\beta_0)/2}\Phi^{\dag}.
\eeq
Conversely, it is impossible to construct a four-fermion
vertex that is invariant under $SU(3)_L \otimes SU(3)_R$
but not $U(1)_A$. This point was emphasized
 by Kobayashi and Maskawa\cite{Kobayashi:1970ji}
and constitutes the basis of their proposal of the six-fermion vertex for 
accounting for the large mass of  $X$ meson, i.e., the $\eta'$.

In fact,
the determinantal terms $\det \Phi$ and 
$\det \Phi^{\dag}$ are
not invariant for the $U(1)_A$ transformation,
although they are invariant under  $SU(3)_L \otimes SU(3)_R$
%because ${\rm det}U=1$ for the special unitary matrix.
% $\det\, Phi^{\dag}$ are not necessarily 
because they are transformed as
\beq
\det \Phi \rightarrow e^{3i(\alpha_0-\beta_0)/2}\det \Phi,\ \ \
 \det \Phi^{\dag} \rightarrow e^{-3i(\alpha_0-\beta_0)/2}
\det \Phi^{\dag},
\eeq
which shows that  ${\cal L}_{_{KMT}}$  is not invariant  unless $\alpha_0=\beta_0$,
i.e., $U(1)_V$ transformation.
Thus, one sees that 
the ${\cal L}_{KMT}$ vertex takes care of the $U(1)_A$ anomaly.
In short,
(i) ${\cal L}_0$ and ${\cal L}_{_S}$ are invariant under 
  $U(3)_L \otimes U(3)_R$
 transformation, while (ii) ${\cal L}_{_{KMT}}$ is invariant
 under  $U(1)_V \otimes SU(3)_L \otimes SU(3)_R$
 transformation but not invariant under $U(1)_A$ transformation.

 ${\cal L}_{_{SB}}$ is the explicit $SU(3)_V$ breaking part with
the current quark masses:
\beq
{\cal L}_{_{SB}}=-\bar{q}{\bfm}q=-\sum _{a=0,3,8}m_a S_a
\eeq
 where
$m_0=(m_u+m_d+m_s)/\sqrt{6},\,  m_3=(m_u-m_d)/2$ and 
$m_8=(m_u+m_d-2m_s)/2\sqrt{3}$ with $S_a=\bar{q}\lambda_aq$.
 If we assume the isospin symmetry,
the mass term is reduced to ${\cal L}_{_{SB}} = -\eps_0S_0-\eps_8S_8$
with $\eps_{0,8}$ being identified with $m_{0,8}$.

 The fact that ${\cal L}_{_{KMT}}$ represents the  $U(1)_A$ anomaly can be
seen in the anomalous divergence of the flavor singlet axial current
\beq
\partial_{\mu}A^{\mu}_5  = 2iN_f g_{_D} ({\rm det}\Phi \, -\,{\rm h.c.})
+ 2i\bar{q}\bfm \gamma_5 q ,
\eeq
with $A_5^{\mu}  =  \bar{u}\gamma^{\mu}\gamma_5u+\bar{d}\gamma^{\mu}\gamma_5d+
\bar{s}\gamma^{\mu}\gamma_5s$.
This equation  is to  be compared with the usual 
anomaly equation\,  \cite{ABJ} written in  terms of the
topological charge density of the gluon field\,
\cite{'tHooft:1976fv,Weinberg:1996kr,Fujikawa:2004cx}, 
\beq
\partial_{\mu} A^{\mu}_5= 2N_f {g^2 \over {32\pi^2}}
 F_{\mu \nu}^a \tilde{F}^{\mu \nu}_a
+ 2i \bar{q}\bfm \gamma_5 q .
\eeq
Thus, one may say that  the effect of the gluon operator
 ${g^2 \over {32\pi^2}} F_{\mu \nu}^a \tilde{F}^{\mu \nu}_a$
is simulated by the determinantal operator 
$i({\rm det}\Phi \, -\,{\rm h.c.}) = -2g_{_D} {\rm Im}({\rm det}\Phi)$
 in the quark sector.
Note that the anomaly term ${\cal L}_{_{KMT}}$ has a dimension 9.
 Away from the chiral limit,
there arise other instanton-induced
 dimension-9 operators that break $U(1)_A$ symmetry but are 
  proportional to the current quark masses\cite{shifman80}.
 The effects of such extra terms are
discussed  in  Ref. \citen{Takizawa:1990ku}.
% and it turns out that the   extra operators do not change the
%qualitative results.  
%Thus  we will not consider them
% in the following analyses for simplicity.

\subsection{Dynamics of Kobayashi-Maskawa-Nambu-Jona-Lasinio Lagrangian}

So far we have discussed  the symmetry properties of each term of the generalized
Nambu-Jona-Lasinio model that
Kobayashi and Maskawa proposed as the low-energy effective model.
What is their dynamical role?
It is easily verified that 
${\cal L}_S$  can be decomposed into
$\sim (\bar{u}\Gamma u)^2\,+\,(\bar{d}\Gamma d)^2\,+\,(\bar{s}\Gamma s)^2$
without the flavor mixing term, like
$(\bar{u}\Gamma u)(\bar{d}\Gamma d)$ with $\Gamma =1$ or $\gamma_5$,
 although there are terms like $(\bar{u}\Gamma d)(\bar{d}\Gamma u)$, and,
 hence does not cause a flavor mixing. On the other hand,
 the KMT  term can cause a flavor mixing 
when the chiral symmetry is dynamically broken; indeed,
  it induces effective 4-fermion vertices such as
$ \la\bar{d}d\ra (\bar{u}u)(\bar{s}s)$ and
$-\la\bar{d}d\ra (\bar{u}i \gamma_5 u)(\bar{s}i \gamma_5 s)$, where
 the former (latter) gives rise to
 a flavor mixing in the scalar (pseudo-scalar) channels.
This flavor mixing in the pseudoscalar channel is found
to be the origin of lifting of the $\eta'$ mass to as high as
1 GeV.
                       
 The vacuum of this model Lagrangian is  determined
 in the self-consistent mean field (SCMF) theory\cite{soken88,Kunihiro:1987bb}
 as is done for the usual NJL model \cite{Nambu:1961tp,Kunihiro:1983ej},
although the six-fermion interaction causes an additional complication in 
the analysis. The  Lagrangian in the SCMF approximation
reads
\beq
 {\cal L}_{_{MFA}} &=& \bar{q}(i\gamma \cdot \partial -\bfM )q
- g_{_S} {\rm Tr}(\phi^{\dagger}\phi)
-2g_{_D}({\rm det}\phi + {\rm c.c.}).
\label{eq:SCMF}
\eeq
Here, the  four-fermion and six-fermion interactions 
 are rewritten in the present approximation
as
\beq
\bar q_iq_i\bar q_jq_j& \rightarrow & \la\bar q_iq_i\ra  
\bar q_jq_j+ \la\bar q_jq_j\ra  \bar q_iq_i
- \la\bar q_iq_i\ra \la \bar q_jq_j\ra ,\nonumber \\  
 \bar q_iq_i \bar q_jq_j\bar q_kq_k& \rightarrow & \sum _{i,j,k;\
{\rm cyclic}}
 \la\bar q_iq_i\ra  \la\bar q_jq_j\ra \bar q_kq_k
 -\ 2 \la\bar q_iq_i\ra \la \bar q_jq_j\ra \la\bar q_kq_k\ra .
\eeq 
The $\phi$ in (\ref{eq:SCMF}) is a diagonal $3\times3$ $c$-number matrix defined 
in terms of the
 quark condensates;
$\phi  =  \langle  \Phi \rangle_0
 \equiv {\rm diag}(\la \bar{u}u \ra, \la \bar{d}d \ra , \la \bar{s}s \ra )$.
 The ``constituent quark mass matrix"
 $\bfM  = {\rm diag} (M_u,M_d,M_s)$ is given in terms of the condensates,
\beq
M_u & =&  m_u - 2g_s \alpha - 2g_{_D} \beta \gamma,\,\nonumber \\
M_d  &=&  m_d - 2g_s \beta - 2g_{_D} \alpha \gamma,\, \nonumber \\
M_s  &=&  m_s - 2g_s \gamma - 2g_{_D} \alpha \beta,
\label{eq:constmass}
\eeq
with $(\alpha, \beta , \gamma)\equiv$
$(\la \bar{u}u \ra, \la \bar{d}d \ra , \la \bar{s}s \ra )$.
 The respective quark  condensate is, in turn, given with $M_i$,
\beq
\la \bar{q}_iq_i \ra =-2N_c\sum_{\vert \bfp\vert<\Lambda}M_i/ {\sqrt{M_i^2 + p^2}}.
\label{eq:vac-con}
\eeq
Here, the three momentum cutoff $\Lambda$ is introduced.
We call Eq. (\ref{eq:vac-con}) together with Eq. (\ref{eq:constmass})
the gap equation because of its resemblance to the gap equation
in the theory of superconductivity\cite{Bardeen:1957kj}.
%For the reduction of the six-fermion vertex, the following 
%identity may be utilized:
%$\det \Phi = {1 \over 3}{\rm Tr}\Phi^3-{1 \over 2}{\rm Tr}
%\Phi^2 {\rm Tr}\Phi + {1 \over 6}({\rm Tr}\Phi)^3$.
 Note that the vacuum condensates
and the constituent quark masses with different 
flavors are all coupled with each other owing to the 
KMT term. To determine these values in terms of the coupling constants,
current quark masses, and the cutoff, one must solve this nonlinear
coupled equation, (\ref{eq:vac-con}) and (\ref{eq:constmass}).
One should verify that the solution of this gap equation really 
gives the true vacuum state by evaluating the vacuum energy or the
effective potential;
\beq
{\cal V}(\phi )   =  i N_c {\rm Tr}\int {d^4p \over (2 \pi)^4} 
{\rm ln}\left(\frac{p \cdot \gamma -M}{p \cdot \gamma - m}\right)
    + g_s(\alpha^2 + \beta^2 +\gamma^2)
+ 4 g_{_D} \alpha \beta \gamma.
\eeq
Here, the first term represents  the  difference in the energy densities of 
 the nonperturbative and perturbative Dirac  seas
  and the second and third terms denote the repulsive interaction energy 
with which the double counting is avoided 
of the attractive interaction energy between quarks through
 the four-fermion and  six-fermion
interactions, respectively.
The stationary condition, $\partial {\cal V}(\phi ) /\partial \phi = 0$ ,
is found to be equivalent to the gap equation, Eq. (\ref{eq:vac-con})
 with Eq. (\ref{eq:constmass}).

Once the vacuum is thus determined, one can discuss the 
meson states as $q$-$\bar{q}$ collective excited states on top of the vacuum.
The residual interactions that develope the 
$q$-$\bar{q}$ collective excitations are the following 
effective four-quark interactions;
\beq
{\cal L}_{res} & = & g_{_S}:{\rm Tr}(\Phi^{\dagger}\Phi): \nonumber \\
&   & +g_{_D}:[{\rm Tr}(\phi\Phi^2)
-{\rm Tr}(\phi\Phi){\rm Tr}\Phi
-{1 \over 2}{\rm Tr}\Phi^2{\rm Tr}\phi+{1 \over 2}
{\rm Tr}\phi({\rm Tr}\Phi)^2 + {\rm h.c.}]: \nonumber \\
&   & + g_{_D}:({\rm det}\Phi + {\rm h.c.}):,
% \nonumber \\
%&   & + ({\rm Fock}\ \ {\rm terms}),
\end{eqnarray}
where the normal ordering is taken with respect to the Fock vacuum of 
${\cal L}_{MFA}$, and we have omitted the Fock terms.

The model can be utilized to describe  the
low-lying pseudoscalar mesons and the  $\eta'$ as
well as 
the scalar mesons and other QCD phenomenology.
As we shall see, the sign of the KMT coupling constant
$g_D$ is found to be negative to reproduce the
mixing properties of the $\eta$-$\eta'$.
This sign assignment is consistent with the identification 
of the KMT term as  the instanton-induced
vertex.\footnote{The sign of $g_D$ is positive for the
two-flavor case, because it is given by 
$\la \bar{s}s\ra\cdot g_D$.}
  First,  we shall 
 show some details on 
how the KMT term can account for 
the $\eta$-$\eta'$ system.

\section{Flavor mixing of $\eta $ and $\eta '$ mesons}

 The relevant interaction in the  $\eta $-$\eta'$ channel 
is found to be
\beq
{\cal L}^{\eta }_{res}={1 \over 2}\sum_{a,b=8,0} :
\eta_aG^{P}_{ab} \eta_b:,
\eeq
where $\eta_a \equiv \bar{q}i\gamma _5\lambda _a q$, and 
 $G^{P}_{ab}$ denotes the coupling constant in the flavor basis,
\beq
{\bfG}^{P} =   \left( \begin{array}{cc}
g_{_S} + {1 \over 3} (  2 \alpha + 2 \beta-\gamma) g_{_D} 
 & - { \sqrt{2} \over 6}(2 \gamma -\alpha -\beta) g_{_D}   \\
- { \sqrt{2} \over 6}(2 \gamma -\alpha -\beta) g_{_D}  
 & g_{_S} - {2 \over 3} ( \alpha + \beta + \gamma ) g_{_D} 
 \end{array} \right).
\label{eq:coupling}
\eeq
The coupling among the modes $\eta_0$ and $\eta_8$  arises both from the
  $SU(3)_V$ breaking and the anomaly terms; 
here, we assume the isospin symmetry ($\alpha=\beta$).
  The effect of large $m_s$ is
 primarily responsible for mixing the octet ($\eta_8$)
 and  singlet ($\eta_0$) modes 
 to make the physical $\eta $ and $\eta '$ mesons.
 One finds  that the $g_{_D}$ contribution is positive in  $G^P_{88}$ but
is negative in $G^P_{00}$ because $g_D\, <0$;
 accordingly,
\beq
G^P_{00}\equiv g_{_S} - {2 \over 3} ( \alpha + \beta + \gamma ) g_{_D} <  
G^P_{88}\equiv g_{_S} + {1 \over 3} (  2 \alpha + 2 \beta-\gamma) g_{_D}.
% < G^P_{33}=G_{\pi }\ \ \ .
\eeq
This inequality implies that the mass of the singlet meson, 
 $\eta_0 = (\bar{u}i \gamma_5 u + \bar{d}i \gamma_5 d
+ \bar{s}i \gamma_5 s)/\sqrt{3}$,
is larger than the octet one,  $\eta_8 =
 (\bar{u}i \gamma_5 u + \bar{d}i \gamma_5
d -2 \bar{s}i \gamma_5 s)/\sqrt{6}$,
 since the binding force is weaker in the singlet channel.
Furthermore, 
noting that the $g_{_D}$ dependence is  strongest in $G^P_{ 00}$, 
one can see that the mass of
 $\eta _{_0}$ is most sensitive to the strength of the
 KMT term.

In the flavor basis, 
$(\bar u i\gamma _5u,\bar d i\gamma _5d,\bar s i\gamma _5s)$,
the coupling constant
matrix ${\bfG}_{\eta \pi^0}$ for  $\eta$ mesons and $\pi^0$  reads
\beq
{\bfG}_{\eta \pi^0}=2\left(\begin{array}{ccc}
g_{_s} & -g_{_D}\gamma & -g_{_D}\beta \\ 
-g_{_D}\gamma & g_{_S} &-g_{_D}\alpha \\
-g_{_D}\beta & -g_{_D}\alpha & g_{_S}
\end{array}\right).
\label{eq:coupling-2}
\eeq
The non-diagonal terms in ${\bfG}_{\eta \pi^0}$ are responsible for
the flavor mixing and,  hence, one sees that  
 not only the strength of the anomaly term $g_{_D}$
but also the quark condensates $(\alpha,\beta$ and $\gamma )$
affect the flavor mixing in the $\pi^{0}$ and $\eta$-$\eta'$ system.

Now, the (un-normalized) propagator of 
the composite ${\bfeta }$-system  in the $SU(3)_V$ basis reads
\beq
{{\bfD}(q^2)} & = & -{\bfG}_{P}^{-1}
 ({1 \over {1 +  {\bfG}_{P}{\bfPi}^{P}(q^2)}}),
\eeq
where
 $({\bfPi }^P)_{ab}$ is the polarization tensor.
The  mixing angle $\theta_{\eta}$ between the $\eta$ and $\eta'$ 
is obtained so that  ${\bfD}^{-1} (q^2)$ is diagonalized as
 \beq
{\bfT}(\theta_{\eta} ){\bfD}^{-1}(q^2)
{\bfT}(\theta_{\eta})^{-1}={\rm diag}(D_{\eta }^{-1}(q^2),
D_{\eta '}^{-1}(q^2)),
\eeq
where the orthogonal matrix $\bfT (\theta_{\eta})$ is given by
\beq
{\bfT}(\theta_{\eta}) =  \left( \begin{array}{cc}
\cos \theta_{\eta}  & \ -\sin \theta_{\eta}  \\
\sin \theta_{\eta}  &  \cos \theta_{\eta}  \end{array} \right).
\eeq
Note that the mixing angle is inevitably energy-dependent in such a
dynamical theory. 

The  $\theta_{\eta}$ is determined through a 
competition between the anomaly (KMT term) and the explicit
 $SU(3)_V$-symmetry breaking ($m_s >> m_{u,d}$);
the former prefers pure $SU(3)_V$ states, i.e., small  $\vert\theta_{\eta}\vert $
 because of the 
 large $u \leftrightarrow s$ and $d \leftrightarrow s$ transitions, while
 the latter
 prefers the mass eigenstates, i.e., large $\vert \theta_{\eta}\vert $.
In fact, 
\begin{itemize}
\item 
if the KMT term is absent ($g_{_D}=0$),
${\bfG}^P=g_{s}\cdot{\bf 1}$ 
and the mass eigenstates are realized, leading to 
the ideal mixing $\theta_{\eta} =-54.75^{\circ}$;
  $\eta = (\bar ui\gamma _5u+ \bar di\gamma _5d)/\sqrt{2}$ 
and $\eta ' = \bar si\gamma _5s$.
On the other hand,
\item
 if $g_D \neq 0$ but $m_s = m_{u,d}$,
 $\theta_{\eta}=0^{\circ}$ and
  the flavor eigenstates are realized as
 $\eta = \eta_8 = (\bar{u}i \gamma_5 u + \bar{d}i \gamma_5
d -2 \bar{s}i \gamma_5 s)/\sqrt{6}$
 and $\eta' =\eta_0 = (\bar{u}i \gamma_5 u + \bar{d}i
\gamma_5 d + \bar{s}i \gamma_5 s)/\sqrt{3}$.
\end{itemize}
 
 An explicit calculation\cite{Kunihiro:1987bb,HKZeit91,Hatsuda:1994pi}
 for the general case gives the results
for the mixing angle at the energy of $\eta$ as 
${\theta_{\eta} (m_{\eta }^2)}  =  -20.9^{\circ}$,
with ${m_{\eta }}  =  486.5$ {\rm MeV} and $ m_{\eta '}=957.5$ {\rm MeV}\,
(fitted).

The parameters of the KM-NJL model to be determined are as follows:
The current quark masses ${\bfm}\, =\, {\rm diag} (m_u,m_d,m_s)$,
the coupling constants $g_{_S}$ and $g_{_D}$,
and the momentum
cutoff $\Lambda$ characterizing the scale of the chiral symmetry
breaking. 
For $m_u$ and $m_d$, we assume the $SU(2)_V$ invariance and
define $\hat{m}$=($m_u$+$m_d$)/2.
These  parameters
are determined\cite{Kunihiro:1987bb,HKZeit91,Hatsuda:1994pi}
 so as to reproduce the four basic quantities
\beq
m_{\pi}=138 {\rm MeV},\,
f_{\pi}=93 {\rm MeV},\, m_{_K}=495.7 {\rm MeV}, \, \, {\rm and}\, \,  
m_{\eta'}=957.5\, {\rm MeV}.
\eeq
We have adopted 5.5 MeV as a value
for $\hat{m}$ at $1$ GeV scale in the following.
Then, the resulting parameter set 
 reads
\beq
\Lambda = 631.4 {\rm MeV},\ \   g_{_S}\Lambda^2=3.666,\ \
g_{_D}\Lambda^5=-9.288,
\ \   m_s=135.7\, {\rm MeV},
\eeq
where we have used a three-momentum cutoff scheme.

\section{The KMT term in scalar meson dynamics}

As was mentioned earlier, the anomaly term of the determinantal form
also gives rise to a flavor mixing in the scalar as well as
in the pseudo-scalar channels.
The scalar mesons may constitute a nonet
as the low-lying pseudo-scalar mesons do,
although some slight difference is surely present
because of the absence of constraints as given by
the axial anomaly in the scalar channel.
Actually, the possible existence of  the low-lying  scalar mesons and
their $SU(3)_V$ nonet scheme were quite 
controversial, although there were some pioneering
works\cite{Delbourgo:1982tv,Hatsuda:1985ey} that 
emphasize the physical significance of the scalar meson, 
particularly  the
$\sigma$ meson in QCD, which has (approximate) chiral symmetry as a 
fundamental property; see also Ref.~\citen{vanBeveren:1986ea}.
The situation has changed completely now since  the two scalar mesons
$\sigma$ and $\kappa$ have been  established experimentally
\cite{Aitala:2000xu,Aitala:2002kr,Ablikim:2004qna,Ablikim:2005ni,DescotesGenon:2006uk}.

The low-lying scalar mesons have attracted renewed interest since
the 1990s when extensive analyses claimed the existence of the $\sigma$ meson pole  
in the complex energy plane of  the   $S$-matrix for the $\pi$-$\pi$ scattering  
in the $I=J=0$ channel\cite{Tornqvist:1995ay,Harada:1995dc,Ishida:1995xx,Ishida:1997wn}.
In these analyses, 
the significance of respecting chiral symmetry, unitarity, and crossing
symmetry was recognized and emphasized 
to reproduce the phase shifts both in the $\sigma$ (s)- and $\rho$ (t)-channels
with a low-mass  $\sigma$ pole\cite{Igi:1998gn}.
One of the most elaborate analyses\cite{Caprini:2005zr}
identifies the $\sigma$ pole
at  $M_{\sigma}=441- i 272$ MeV.
 The existence of such low-lying scalar mesons
can be a puzzle in QCD \cite{Kunihiro:2007yw,Close:2002zu}.
In the nonrelativistic constituent quark model\cite{CLOSE},
the meson with the quantum number $J^{PC}=0^{++}$  is in the $^3P_0$
 state, which normally implies that 
the mass lies in the region from $1.2$ to $1.6$ GeV.
Several mechanisms have been proposed to lower the mass with 
an amount as large as $600$ -- $800$ MeV; see Ref.~\citen{Kunihiro:2007yw}
 for the issues concerning the low-lying scalar mesons.
 A first idea was a diquark-anti-diquark (or tetraquarks) structure
proposed by Jaffe\cite{Jaffe:1976ig}, who showed that
the color magnetic interaction between the diquark and
the anti-diquark gives a sufficiently large attraction to 
decrease the masses of
the scalar mesons to approximately  600 MeV.
 Another time-honored idea is attributable to Nambu\cite{Nambu:1961tp}, 
in which the smallness of the mass is attributed 
 to  the possible collective nature of  the scalar mesons as possessed
by the pion.
 It is well known  that the scalar meson appears
as a consequence of the  chiral symmetry and its dynamical
breaking as the pion does, and the mass of the 
sigma satisfies the Nambu relation\cite{Nambu:1984ph},
$m_{\sigma}=2 M_f$, with $M_f$ being the dynamically generated fermion 
(quark) mass, which should be valid within any Nambu-Jona-Lasinio
type model. If we put $M_f=300$ MeV, $m_{\sigma}$ becomes 600 MeV,
in fairly good agreement with the experiment. 
It is shown that this feature essentially persists even 
when the $U(1)_A$ anomaly term is incorporated
although there arises a small but sizable flavor 
mixing between the $\sigma \sim (\bar{u}u+\bar{d}d)/\sqrt{2}$ and 
$f_0 \sim \bar{s}s$\, \cite{Kunihiro:1987bb,HKZeit91}; 
see also some subsequent works \cite{Dmitrasinovic:1996fi,Celenza:1997it,Umekawa:2004js}.
The wave functions of scalar mesons should  also have 
components of  meson-resonance states as these states are seen
through the $\pi$-$\pi$ or $\pi$-$K$ scattering.

Although the reality should be that the wave functions of 
these mesons are linear combinations of these components,
the most popular idea is the tetraquarks
\cite{Jaffe:1976ig,Black:1999yz,Maiani:2004uc,Fariborz:2007ai}.
In this scheme, the $SU(3)$ nonet structure is composed
of  the quark content as follows;
$\sigma=[ud][\bar{u}\bar{d}]$,\,
$\kappa^0 = [su][\bar{u}\bar{d}]$,\,
$\kappa^{-} = [sd][\bar{u}\bar{d}]$,\,
$f_0=([su][\bar{s}\bar{u}]+[sd][\bar{s}\bar{d}])/\sqrt{2}$,\,
$a_0^{+}=[su][\bar{s}\bar{d}]$, and
other members, $\kappa^{+,-},\, a_0^{0,-}$, are
constructed in a similar manner. 

The merit of the tetraquark scheme lies in the fact \cite{Jaffe:1976ig}
that it can naturally identify the nonet scheme in the 
scalar mesons of the masses of less than 1 GeV and also explain the  multiplet scheme
 that has an inverted form of the vector-meson nonet, i.e.,
 the $\rho$, $\omega$, $K^{*}$, and $\phi$.
However, there are at least two problems to be clarified to establish this scheme
\cite{Hooft:2008we}:
(i)~
If the flavor mixing is
ideal as given above, the $f_0\rightarrow 2\pi$ coupling vanishes
in contrast to that in the experiment, 
and the $a_0\rightarrow \eta \pi$ coupling is too large
to be consistent with the experimental data.\,
(ii)~As is mentioned above, there should be a mixing between the tetraquark
and $q\bar{q}$ states more or less to make the physical states, which 
may possibly imply the existence of the scalar mesons mainly composed 
of $q\bar{q}$ with a small mixture of the tetraquark states.
Although there is work on these problems\cite{Teshima:2001dw},
 several researchers \cite{Fariborz:2008bd,Hooft:2008we} 
have recently  shown that the KMT term in the scalar meson dynamics
can nicely  resolve these problems. 
By making a Fiertz transformation, one can see that 
the KMT term contains
a tetraquark-$q\bar{q}$ coupling,
\beq
{\cal L}_{4q-2q}=G_{4q-2q} \Tr\big(\tilde{S}S\big),
\label{eq:Fiertz}
\eeq
where
$\tilde{S}_{ij}=[\bar{d}]_i[d]_j$ and
$S_{ij}=\bar{q}_jq_i$ with 
$[d]_{i\alpha}=\eps_{ijk}\eps_{\alpha\beta\gamma}\bar{q}_c^{j\beta}\gamma_5q^{k\gamma}$
being the spin-$0$ diquark operator. The Latin and Greek indices 
denote flavor and color state, respectively, and 
$q_c$ is the charge conjugate of the quark field.
This form of the vertex was first considered in Ref. \citen{Black:1999yz}
in the context of the scalar meson mixing.\footnote{
We remark that the vertex in the form of Eq. (\ref{eq:Fiertz}) was
also considered in the context of the color superconductivity 
in dense quark matter\cite{Alford:1998mk,Rapp:1999qa,Steiner:2005jm,Yamamoto:2007ah};
see \S 6.}
Although the coupling constant $G_{4q{\rm -}2q}$ is in principle given
by the KMT coupling $g_{_D}$, the phenomenological analysis
indicates \cite{Black:1999yz} that
$\vert G_{4q{\rm -}2q}\vert\simeq 0.6$ GeV$^2$.
See Refs. \citen{Fariborz:2008bd} and \citen{Hooft:2008we} for the details
of the roles of the KMT vertex in the tetraquark-$q\bar{q}$
mixing and the resulting phenomenology for the scalar meson dynamics.

\section{Other phenomenology with the KMT term}

It has been shown that  the KM-NJL model well describes the vacuum properties
 related to chiral symmetry and its spontaneous breaking
 including their flavor dependence.\cite{soken88,Kunihiro:1987bb,Hatsuda:1994pi} 
 We have seen that the model gives a systematic description of the 
 low-energy hadrons in the pseudo-scalar  and scalar channels. 
The model can be a good starting point even for 
 the octet and decuplet {\em baryons};
\cite{Kunihiro:1990ts,Hatsuda:1994pi}\footnote{ 
See also Ref. \citen{Oka:1989ud,Takeuchi:1990qj,Oka:1990vx},
in which the role of the instanton-induced interaction is examined
for the existence or nonexistence of the $H$-dibaryon\cite{Jaffe:1976yi}.}
 one can  use the vertex 
 to analyze the possible violation of the  OZI rule in the baryon sector.
  A fundamental reason for such successes of the KM-NJL model lies in the 
 fact that the  model can be regarded as a field theoretic version of 
 the constituent quark model under the identification of the constituent
 quark masses with those generated dynamically by the 
chiral symmetry breaking: The new ingredients of the KM-NJL model beyond the
 conventional constituent quark model are (i) it gives a  
 self-consistent   description of the {\em vacuum} and the excited states
 ({\em hadrons}),
 and  (ii)  the model properly 
 takes into account  the {\em collective} nature of the vacuum and 
the mesons. These points are emphasized in Ref.~\citen{Hatsuda:1994pi}.
The recent development of the phenomenology based on the KMT term may be
seen in Ref.~\citen{Dmitrasinovic:2006de} and the references cited therein.

\section{Application to finite temperature and density systems}

The role  of the $U(1)_A$ anomaly at finite temperature $T$
and/or baryon density$\rho_B$ or the chemical potential $\mu$ is
 a big issue, 
and there are  many studies on this problem; see, for example, 
Ref.~\citen{Gross:1980br,Pisarski:1983ms,Schafer:1996wv}. 
As an effective model embodying the $U(1)_A$ anomaly,
the KM-NJL model is also applied to the finite temperature and density.
Here, we will pick  some topics 
on these subjects.

\subsection{Phase diagram}

  In Ref.~\citen{Kunihiro:1989my},
the quark condensates and meson excitations in the hadronic 
phase at finite temperature $T$
are investigated by the present author 
in the KM-NJL model for the first time; 
see also Refs.~\citen{soken88} and \citen{Kunihiro:1991hp}. It was shown that 
the order of the chiral transition is of {\em crossover} against an expectation
that the cubic term owing to the KMT term would lead to a first-order phase transition.
 This is because the explicit symmetry terms owing to the 
current quark masses, particularly that of the strange quark, are so large
that the order of the phase transition becomes smooth.
This is the result within the mean-field approximation.
Thus, it would be intriguing to apply the functional renormalization
group to explore the chiral phase transition with the KMT term and the
explicit breaking with the current quark masses.

The mean field theory at finite $T$ and $\mu$ goes much the same way as
 that at zero temperature discussed in a previous section:
The vacuum expectation 
 value $\la O\ra$ is replaced by the statistical average $\lla  O \rra$.
Then,  the quark condensates as the variational parameters 
 are now $T$- and $\mu$-dependent;
$
\lla\bar uu\rra\equiv \tilde{\alpha},\ \  \lla\bar dd\rra \equiv 
\tilde{\beta},\ \  \lla\bar
ss\rra \equiv \tilde{\gamma}.$
Thus, the Hamiltonian  to be used  in this approximation  has the same 
form as that at zero temperature,
with the quark mass matrix ${\bfM}={\rm diag}(M_u,M_d,M_s)$ given
by Eq. (\ref{eq:constmass}) but now
being $T$- and $\mu$- dependent:
$M_u  =  m_u - 2g_s \tilde{\alpha} - 2g_{_D} \tilde{\beta} \tilde{\gamma}$,\,
$M_d  =  m_d - 2g_s \tilde{\beta} - 2g_{_D} \tilde{\alpha} \tilde{\gamma},$\,
and $M_s  =  m_s - 2g_s \tilde{\gamma} - 2g_{_D} \tilde{\alpha} \tilde{\beta}$.\,
 We note again that the contribution of the anomaly term to 
the constituent quark masses is dependent on the condensates of 
other flavors.  It indicates that a change, say, in $\lla\bar uu\rra $ 
causes a change in $M_s$ and accordingly in $\lla\bar ss\rra $,
and {\it vice versa}.  Thus, 
the properties of the strange quark can change 
even in the  matter composed of the $u$ and $d$ quarks,
i.e., nuclear matter.

The thermodynamical potential in the mean-field approximation
 can be readily calculated with 
\beq
  K_{MFA}= H_{MFA}-\sum _{i=u,d,s} \mu _i N_i.
\eeq
The result is 
\beq
\Omega_{MFA}(\tilde{\alpha} ,\, \tilde{\beta} ,\, \tilde{\gamma} ) &=&{\cal
V}(\tilde{\alpha} ,\tilde{\beta} ,\tilde{\gamma} )\cdot V-2N_cT\sum _{i=u,d,s,\vert p\vert<\Lambda}
[\ln\lbrace 1+\exp (-e^{(-)}_{i{\bfp}}/T)\rbrace \nonumber \\ 
\ \  &\ \ \ & +\ln \lbrace 1+\exp (-e^{(+)}_{i{\bfp}}/T)\rbrace ],
\eeq
where $e^{(\pm)}_{i{\bfp}}=E_{i{\bfp}}\pm \mu _i$ with 
$E_{i{\bfp}}=\sqrt {M_i ^2+{\bfp} ^2}$ and $V$ being the volume of
the system,  and 
\beq
{\cal V}(\tilde{\alpha},\tilde{\beta},\tilde{\gamma})=-2N_c\sum _{i=u,d,s}\int ^{\Lambda}
{{d{\bfp}}\over {(2\pi )^3}}
E_{i{\bfp}}+\lbrace  g_s(\tilde{\alpha} ^2+\tilde{\beta}^2+\tilde{\gamma}  ^2)+
4g_D\,\tilde{\alpha} 
\tilde{\beta}\tilde{\gamma} \rbrace 
\eeq
 is the vacuum energy term, which has the same form as the  effective 
potential at $T=0$, although the condensates are now 
 temperature-dependent.
We have neglected the  constant term 
$\Omega (0, 0, 0)$, which is irrelevant for the following argument.
The equilibrium state can be determined as the point where the
thermodynamical potential takes the minimum with 
$\tilde{\alpha},\,\tilde{\beta} $, and
$\tilde{\gamma} $ as the variational parameters:
\beq
{{\partial \Omega _{MFA}} \over {\partial Q_i}}=0,\ \ \ \ \ \ \
 (Q_i=\tilde{\alpha},\tilde{\beta},\tilde{\gamma})
\eeq
 which ensures that
 the condensates assumed 
are the statistical averages calculated with the corresponding
mean-field Hamiltonian.

A numerical calculation \cite{Kunihiro:1989my,Kunihiro:1991hp}
 shows that the thermodynamical potential 
$\Omega (\tilde{\alpha}, \tilde{\beta}, \tilde{\gamma} )$ as a function 
of the condensates has
  only one minimum  for all the temperatures  with vanishing
chemical potentials.  This  implies that the phase transition described
in this model is a smooth one, or a crossover\cite{Kunihiro:1989my}. 
However,
the transition turns out to 
  be a first-order one  at finite density with low
temperatures ($T\leq 29$ MeV $\equiv T_{c1}$) like a liquid-gas phase transition:
 At
$T=0$, a chirally restored phase with a high density ($\rho _B\sim
6\rho _{_0}$, $\rho _{_0}=.17\, {\rm fm}^{-3}$) coexists with a chirally
broken phase with a small density. As  $T$ increases, the difference
in the densities of the coexisting phase becomes smaller and
smaller; then, at $T=T_{c1}$, the phase transition ceases to be a 
first-order one. For $T > T_{c1}$,
 the phase transition is a smooth one, and the pressure
increases monotonically as the density is increased
\cite{Kunihiro:1989my,Kunihiro:1991hp}.

\subsection{The mesonic excited states; effective restoration of U(1)$_{\rm A}$
symmetry}
% as seen in the $\eta$-$\eta'$ system at finite temperature}

Now let us proceed to the examination of the meson states as
collective excitations in the system.

%\subsubsection{Dispersion equations}

The collective excitations are generated primarily 
by the four-fermion interactions;  the  effective coupling constants 
$G_{\alpha }$  in the various channels are tabulated
 in Table 3.1 in Ref.~\citen{Hatsuda:1994pi}.
For the $\eta$-$\eta'$ channel,
the coupling matrix is given 
in Eq.~(\ref{eq:coupling}) or Eq.~(\ref{eq:coupling-2}).
 Note, however, that the condensates appearing there
are now $T$- and $\mu _i$-dependent\cite{Kunihiro:1989my}.
 Furthermore, since the condensates are multiplied by $g_{_D}$ there,
 the net effects of the anomaly as manifested in the mixing 
properties of the $\eta $ and $\eta '$ mesons
should become smaller when $T$ is increased.
 This means that the chiral restoration effectively causes a partial
restoration of the $U(1)_A$ symmetry\cite{Kunihiro:1989my}.

The information of collective excitations is all contained in the
corresponding retarded Green's functions or the response functions
given by
\beq
R_{\alpha \beta}(\omega,{\bfq})=-i\int \frac{d^4x}{(2\pi)^4}
e^{-iq\cdot x}\theta(t)\lla[ {O}_{K \alpha}(t,{\bfx}),
 {O}_{K \beta}(0,{\bfzero})]_{-}\rra ,
\eeq
where 
\beq
 {O}_{K \alpha}(t, {\bfx})=\bar{q}_{_K}(t,{\bfx})\Gamma _{\alpha}
q_{_K}(t,{\bfx})-\lla\bar{q}_{_K}(t,{\bfx})\Gamma _{\alpha}
q_{_K}(t,{\bfx})\rra,
\eeq
with $q_K(t, {\bfx})=\exp(-i {K}t)q(0,{\bfx})\exp (i {K}t)$ 
being the real-time  operator.
Here, $\Gamma _{\alpha}(\Gamma _{\beta})$ denotes  a product of  Dirac and 
Gell-Mann matrices that specifies the
quantum numbers of the collective modes; for example, $\Gamma _{\alpha
}=\Gamma _{\beta}=i\gamma _5\lambda _{4\pm i5}$ for kaons $K^{\pm}$.
For the sigma mesons and $\eta $ and $\eta '$
mesons, the response functions become  matrices owing to  the
octet-singlet couplings. The
poles of the response function (or the determinant of the response
functions  for sigma
and $\eta $ mesons) give the dispersion relations $\omega _{\alpha }
=\omega _{\alpha }({\bfq})$ of the mode.
To evaluate the response function, one may use the {\it imaginary-time}
formalism.

It was shown \cite{Kunihiro:1989my} that 
the $\eta$ and $\eta^\prime$ mesons change
their nature owing to both the temperature dependence of
 the quark condensates and the possible decrease
in the KMT coupling constant $g_{_D}$ with $T$.
The coupling constant $g_{_D}$ of the KMT term may be dependent 
on temperature and baryon chemical potential because the instanton density is
dependent on them\cite{Gross:1980br,Schafer:1996wv}.
When such a possible temperature dependence is considered,
the mixing angle $\theta_{\eta}$ can also be $T$ dependent,
 and $\theta_{\eta}$  increases in the absolute value and
the mixing between the $\eta$ and $\eta'$ approaches the ideal 
one.
Although the $\eta_0$ component in the physical $\eta'$ 
decreases as $T$ is increased,
the $\eta^{\prime}$  mass decreases gradually with increasing
$T$, because 
the $\eta_0$ tends to acquire the nature
of the ninth Nambu-Goldstone boson of
the $SU(3)_L \otimes SU(3)_R \otimes U(1)_A$ symmetry 
and decreases its mass rapidly. 
 This tendency is also observed with an explicit use of
the instanton-induced interaction\cite{Schafer:1996hv,Schafer:1996wv}.
This is an effective ``restoration'' of $U(1)_A$ anomaly
at finite temperature as seen in the $\eta$-$\eta'$ spectrum,
 which was first suggested by Pisarski and Wilczek\cite{Pisarski:1983ms}
 using a linear $\sigma$ model with a determinant term in the chiral limit.
Such an anomalous decrease in the $\eta^{\prime}$ mass
might have been observed in 
the relativistic heavy ion collisions at RHIC\cite{csorge}.

\subsection{Possible temperature and density dependence of 
the KMT coupling $g_{_D}$}

The temperature dependence of  $g_{_D}$ can be deduced
by utilizing  the lattice data, as was done by
 Fukushima, Ohnishi, and Ohta\cite{Fukushima:2001hr}.
The finite density case is examined in
Ref.~\citen{Costa:2002gk}.
The possibility of the effective restoration 
of the chiral $U(1)_A$ anomaly in finite nuclei is discussed
in Ref.~\citen{Nagahiro:2004qz} where
 a possibility to create  bound states of the $\eta'(958)$ meson in nuclei 
is examined. Further developments in this direction may be found in
Refs.~\citen{Bass:2005hn} and \citen{Nagahiro:2006dr}.
The density dependence of the coupling constant
$g_{_D}$ has also been considered \cite{Chen:2009gv}
for exploring whether
the QCD critical point  suggested in some effective models\cite{Asakawa:1989bq}
can actually be absent as is shown in some lattice simulation\cite{deForcrand:2006pv}.
 
\subsection{Incorporation of color superconductivity;\,
the U(1)$_{\rm A}$ anomaly versus vector interaction}

At extremely high density with  moderate temperature,
various forms of color superconductivity may occur;
see Refs.~\citen{Buballa:2003qv} and \citen{Alford:2007xm} for the recent reviews.
In the three-flavor case in the chiral limit,
the  most symmetric pairing can be realized, which is 
called the color-flavor locked (CFL) phase\cite{Alford:1998mk}.
On the basis of the pattern of the symmetry breaking in the CFL phase,
Schafer and Wilczek \cite{Schafer:1998ef} suggested a hadron-quark 
continuity,
i.e., the transition from hadron to quark matter can be smooth.
As was stated before in the context of the tetraquark structure 
of the scalar mesons,
a Fiertz transformation of the KMT term gives a 
tetraquark-$q\bar{q}$ coupling that breaks the $U(1)_A$ 
symmetry\cite{Alford:1998mk,Rapp:1999qa,Steiner:2005jm,Yamamoto:2007ah}.
It was speculated \cite{Yamamoto:2007ah} that the existence of
such a vertex would give rise to another
QCD critical point in the low-temperature region
and can be essential to realize the hadron-quark continuity\cite{Schafer:1998ef}.
Here, we remark that the possible
 existence of multiple critical points in QCD phase diagram
was first shown in Ref.~\citen{Kitazawa:2002bc} where
the vector interaction plays the essential role;
see also Ref.~\citen{Zhang:2008wx}.
The combined effects of the KMT term as well as the vector interaction 
on the QCD phase diagram have been recently
examined in Ref.~\citen{Zhang:2009mk},
 together with the charge neutrality and the beta equilibrium
constraints.

\section{Concluding remarks}

In the present article, we have described the significance 
of the work by Kobayashi and Maskawa in 1970\cite{Kobayashi:1970ji}, which
introduced the determinantal six-quark interaction
 to account for the large mass
of  $X$ meson, which is now called the $\eta'$,
although the determinantal term is often called the
't~Hooft vertex because it can be derived from
the instanton-induced interaction.
Then, we proposed to call the determinantal six-quark
interaction the Kobayashi-Maskawa-'t~Hooft (KMT) term. 
We have also indicated that  the effective Lagrangian suggested in the
Kobayashi-Maskawa paper in 1970 is actually 
the generalized Nambu-Jona-Lasinio model with the KMT term,
which has now been
widely used both in QCD phenomenology and
in the extensive study of the condensed matter physics of QCD
at finite temperature and density.
Some focus was put on the recent active studies on
 the scalar meson dynamics with diquark correlations, which,
in turn, can give rise to color superconductivity in high-density quark matter.

Finally, I wish to tell my personal experiences 
with Professor Maskawa and Professor Kobayashi, particularly that related to
their work\cite{Kobayashi:1970ji}.
I gave a talk on my work on the generalized NJL model with the KMT term
 at a  workshop held 
at the Research Institute for Fundamental Physics (RIFP)\footnote{
The former and original  name of the Yukawa Institute for Theoretical Physics
(YITP).} from November 4 to 6, 1987. 
My talk\cite{soken88} entitled ``An Effective Theory of QCD --- SU(3)-Nambu-Jona-Lasinio
Model Incorporating the Anomaly Term ---'' 
 consisted of a part that corresponds to
\S\S 2 and 3 in the present article and a sketch on the application
of the model  to finite temperature and density\cite{Kunihiro:1989my,Kunihiro:1991hp},
together with a discussion on the vector mesons.
The proceedings of the meeting were published in 
a Japanese  journal called Soryushiron Kenkyu\cite{soken88}
 in July 1988.
I became aware of the work by 
Kobayashi and Maskawa\cite{Kobayashi:1970ji}
and the subsequent work by Kobayashi-Kondo-Maskawa\cite{Kobayashi:1971qz}
after the domestic meeting,
and therefore had not cited their papers in the proceedings\cite{soken88}.
Although I do not remember exactly, 
my collaborator, Tetsuo Hatsuda, 
 might possibly have told me of their papers;
although he was involved in other projects \cite{Hatsuda:1987tu} at KEK  as 
a postdoc there,
he kindly helped me by my request after the  meeting at RIFP
to rapidly finalize the paper\cite{Kunihiro:1987bb},
which was submitted at the very end of 1987.
I  remember
that in a certain  meeting held at RIFP
in July or August, 1988, Professor Maskawa sat  next to me
and, to my surprise,  talked to me, and we chatted (in Japanese, of course),
 roughly  as follows:\\
Maskawa:\, Are you the author who did the analysis of the determinantal
interaction that is reported in the latest Soken\footnote{An abbreviation
of Soryushiron Kenkyu in Japanese.}?\\
T.~K.:\, Yes, I am.\\
Maskawa:\, Some years ago, we made an analysis of the $\eta'$
and concluded that there must be a six-fermion determinantal
interaction for describing the $\eta'$.\\
T.~K.:\, I know of your papers. \\
This brief chat with Professor Maskawa was a significant event and
of great encouragement to me, since I had been feeling that my works were
 not fully appreciated, although
I had a strong confidence in my work, particularly
in that presented in the meeting at RIFP, as is described in the
Introduction of the proceedings\cite{soken88}.
 Professor Maskawa happened to
be the director of YITP when I earned a position there in 2000.
Unfortunately, I failed to ask him whether he remembered the event described
above.

 Professor Kobayashi was the supervisor of our
exercise class on electromagnetism when I was a
student at Kyoto University.  He was still young and 
an assistant professor then.
Tetsuo Matsui, a former classmate of mine
and now at the University of Tokyo, 
was brave enough to ask Professor Kobayashi
to  tutor our group who planned to read the textbook
on quantum mechanics of Dirac. He kindly accepted our request.
That was from 1972 to 1973,
which means that although 
he might have been busy
 developing the Kobayashi-Maskawa theory on the 
$CP$ violation\cite{Kobayashi:1973fv},
he was  kind enough to take the time to supervise  us in reading a textbook on
quantum mechanics. 
I was  fortunate that Professor Kobayashi was
also in charge of  the seminar on elementary particle physics
when I entered the graduate school of Kyoto University.
Moreover,  he chose the paper by Nambu-Jona-Lasinio\cite{Nambu:1961tp}
as one of the papers that we were to read and report in the
course.
I now appreciate 
how much  my career is owed to these two great physicists. 
It is a great honor and pleasure for me
to contribute to this special issue to celebrate the
Nobel Prize awarded to Professor Maskawa and Professor Kobayashi
by writing an article on the subject through which I 
crossed paths with them.

\section*{Acknowledgements}

First of all, the author thanks 
Professor Kugo and Professor Onogi  for inviting  him 
to write this article on the KMT term.
The author acknowledges Akira Ohnishi and Kenji Fukushima for their
 careful reading of the manuscript and helpful comments.
This work was partially supported by a
Grant-in-Aid for Scientific Research by the Ministry of Education,
Culture, Sports, Science and Technology (MEXT) of Japan (No.
20540265),
 by the Yukawa International Program for Quark-Hadron Sciences, and by the
Grant-in-Aid for the global COE program ``The Next Generation of
Physics, Spun from Universality and Emergence'' from MEXT.

%\appendix
%\section{First Appendix} %Empty argument \section{} yields `Appendix'. 
%
%\section{Second Appendix}

\end{document}